\documentclass{iaus}
\usepackage{graphicx}

\title[NStED-ETSS] 
{The NStED Exoplanet Transit Survey Service}

\author[K. von Braun et al.]   
{K. von Braun$^{1,4,5}$,
M. Abajian$^{2,4}$,
B. Ali$^{2,4}$,
R. Baker$^{2,4}$,
G.B. Berriman$^{1,2,4}$,
N-M. Chiu$^{2,4}$,
D.R. Ciardi$^{1,4}$,
J. Good$^{2,4}$,
S.R. Kane$^{1,4}$,
A.C. Laity$^{2,4}$,
D.L. McElroy$^{2,4}$,
S. Monkewitz$^{2,4}$,
A.N. Payne$^{1,2,4}$,
S. Ramirez$^{2,4}$,
M. Schmitz$^{2,4}$,
J.R. Stauffer$^{3,4}$,
P.L. Wyatt$^{1,2,4}$,
\and A. Zhang$^{2,4}$}

\affiliation{$^1$ Michelson Science Center;
$^2$ Infrared Processing and Analysis Center;
$^3$ Spitzer Science Center;
$^4$ California Institute of Technology;
$^5$ email: {\tt kaspar@caltech.edu}}

\pubyear{2008}
\volume{253}  
\pagerange{1--4}
\jname{Transiting Exoplanets}
\editors{Editor, eds.}
\begin{document}

\maketitle


\begin{abstract}
The NASA Star and Exoplanet Database (NStED) is a general purpose stellar
archive with the aim of providing support for NASA's planet finding and
characterization goals, stellar astrophysics, and the planning of NASA and
other space missions. There are two principal components of NStED: a database
of (currently) 140,000 nearby stars and exoplanet-hosting stars, and an
archive dedicated to high-precision photometric surveys for transiting
exoplanets.  We present a summary of the latter component: the NStED Exoplanet
Transit Survey Service (NStED-ETSS), along with its content, functionality,
tools, and user interface. NStED-ETSS currently serves data from the TrES
Survey of the Kepler Field as well as dedicated photometric surveys of four
stellar clusters. ÊNStED-ETSS aims to serve both the surveys and the broader
astronomical community by archiving these data and making them available in a
homogeneous format.  Examples of usability of ETSS include investigation of
any time-variable phenomena in data sets not studied by the original survey
team, application of different techniques or algorithms for planet transit
detections, combination of data from different surveys for given objects,
statistical studies, etc. NStED-ETSS can be accessed at
\tt{http://nsted.ipac.caltech.edu}.

\keywords{astronomical data bases: miscellaneous; catalogs; surveys; time;
stars: variables; planetary systems}
\end{abstract}


\firstsection 


\section{Specific Goals of NStED-ETSS}\label{goals}

The purpose of NStED-ETSS is to make available to the astronomical community
time-series data (i.e., light curves) of planet transit studies and other
variability surveys in a homogeneous format, along with tools for data
analysis and manipulation. The principal goals of NStED-ETSS include the
following:
\begin{itemize}
\item Provide access to support data for ground-based and space-based missions.
\item Allow the development of different or improved algorithms for transit
detection or variability classification on complete existing survey data sets;
for instance, to enable the detection of planets previously missed in the
original study.
\item Extend the time baseline for transit studies by using data sets
containing the same stars, leading to increased detection efficiency, results
of increased statistical significance, enhanced potential to conduct transit
timing studies, etc.
\item Enable improved understanding of false positivies encountered in transit
surveys.
\item Provide access to a wealth of other astrophysical results and ancillary
science not pursued in the original survey, such as studies of eclipsing
binary and other variable stars or variability phenomena, stellar atmospheres
(rotation, flares, spots, etc.), asteroseismology and intrinsic stellar
variability, as well as serendipitous discoveries such as photometric
behaviors of supernovae progenitors, etc.
\end{itemize}


\section{ETSS Organization \& Visualization}\label{organization}

Each data set contained in ETSS features a {\it master file} and many {\it
light curve files}. Tools enable the user to visualize the data and perform
manipulation and analysis tasks. 

The master file provides basic properties of the data set as a whole as well
as global parameters about each individual light curve file. Through the NStED
infrastructure, one may thus use the master file to search the data set by
parameters such as unique identifiers, celestial coordinates, static
photometry parameters (single-epoch magnitudes), variability filter(s),
Heliocentric Julian Dates (HJD) of the first and last data points, number of
observational epochs, rms dispersion about the median magnitude, existence
(and frequency) of photometric outliers, $\chi^2$ about the median magnitude,
cross identification between different magnitudes, etc.

Each light curve file is associated to a unique identifier and features a
header summarizing global information about the light curve, as well as the
column-delimited photometry data, generally in the format HJD, magnitude,
uncertainty. Thus, it is flexible and readable with all computer operating
systems and can easily be translated to other formats such as VO, binary fits
tables, etc.

\begin{figure}
\begin{center}
 \includegraphics[scale=0.8]{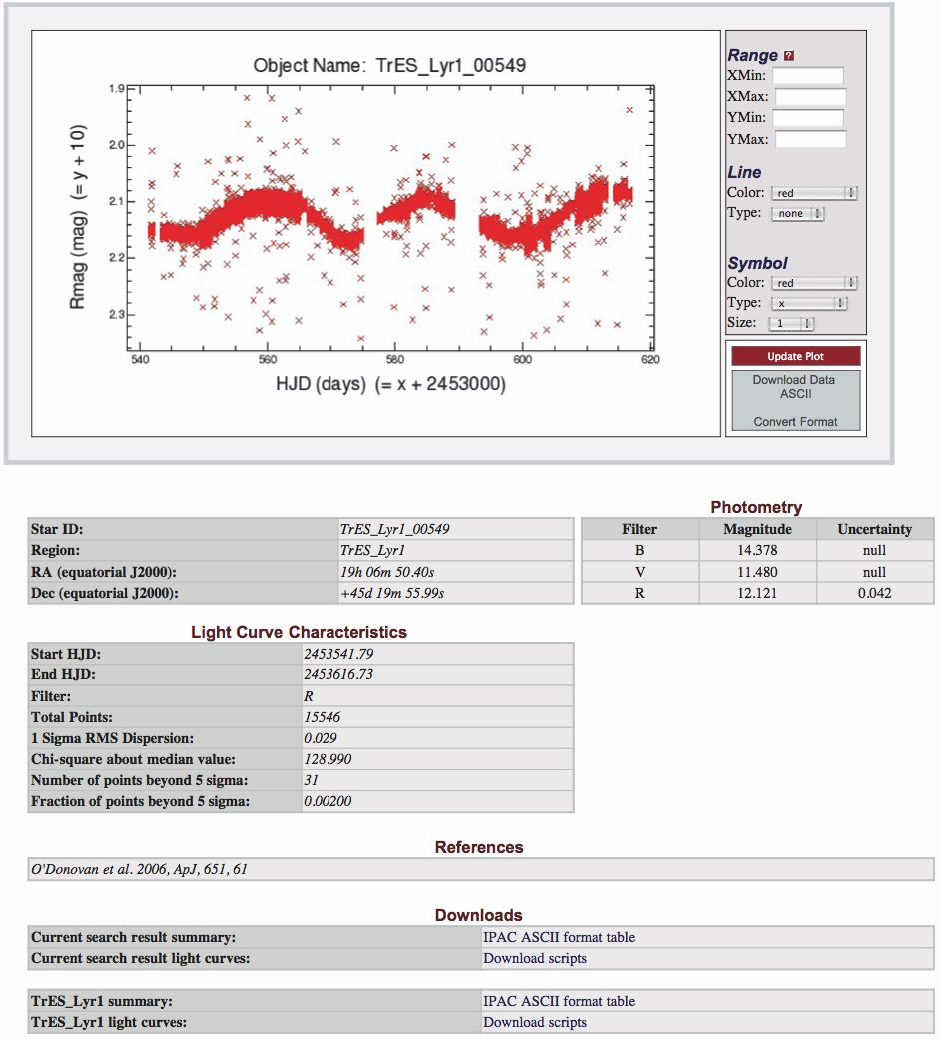} 
 \caption{ETSS Detail Page: featured are an interactive light curve viewer
 (mag vs HJD), summary of light curve characteristics, direct links to ascii
 light curve, cross-identified stars (if applicable), summary table, download
 scripts, and the data set summary (master file). For details, see
 \S\ref{organization}.}  \label{fig1}
\end{center}
\end{figure}

Fig. \ref{fig1} shows an example of data visualization found on the NStED-ETSS
website, complete with light curve characteristics, data set reference, and
links to the associated  files and download scripts. The data in this
plot are taken from the TrES-Lyr1 data set, donated by F. T. O'Donovan and
described in \cite{ocm06}.

\section{ETSS Holdings and Future Data Sets}\label{holdings}

\begin{figure}
\begin{center}
 \includegraphics[width=\textwidth]{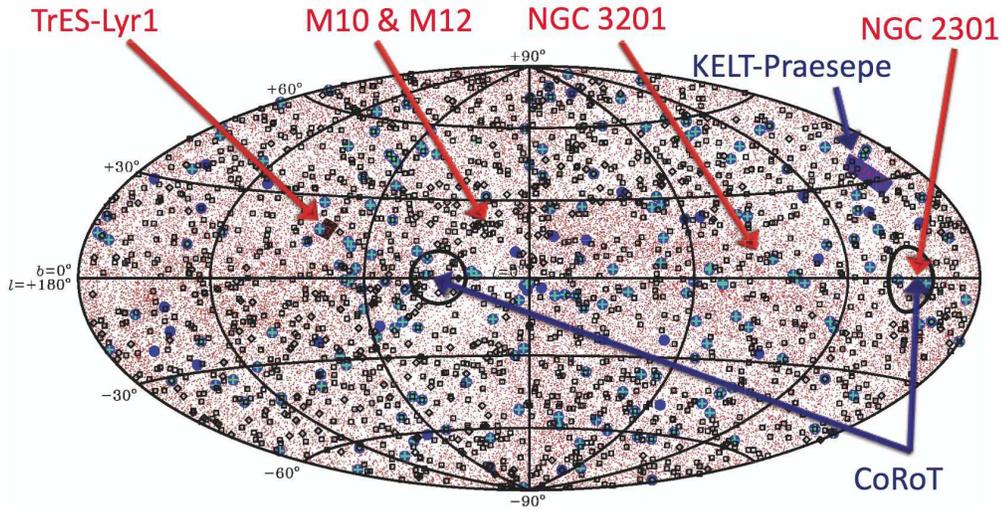} 
 \caption{NStED-ETSS Contents: Aitoff projection with the locations of the
 current and future survey data sets. For an explanation of points and
 squares, please see companion paper on the NStED Stellar Service (Ramirez et
 al. 2008, this volume). Shown are the locations of the three globular
 clusters (M10, M12, NGC 3201), the open cluster (NGC 2301), and the TrES-Lyr1
 field, all of which are contained in the current version of NStED-ETSS. Also
 shown are future data sets: two CoRoT fields (circles along the Galactic
 plane), and the KELT-Praesepe data set; both are expected later this
 year. For details, see \S\ref{holdings}.}  \label{fig2}
\end{center}
\end{figure}

Fig. \ref{fig2} shows the current and some near-term future data sets featured
in ETSS. TrES-Lyr1, the TrES network planet transit survey of a field in Lyra,
described in \cite{ocm06}, contains $\sim$ 26,000 stars with 15,500
observation epochs over 75 nights in the $R$ and $r$ filters. The data sets on
the globular clusters (GCs) M10 and M12 contain 44,000 and 32,000 stars,
respectively, with $\sim$ 50 observational epochs in both $V$ and $I$ over a
500-night timespan (\cite{bmc02}). The data set on the GC NGC 3201 features
$\sim$ 59,000 stars with 120 epochs in each $V$ and $I$ over the course of 700
nights (\cite{bm01, bm02}). NGC 2301 is an open cluster and its data set
contains 150 epochs in $R$ on 4,000 stars over 14 nights (\cite{hvt05,the05}).

Data sets that are expected later in 2008 include the KELT-Praesepe field,
described in \cite{ppd07} and \cite{psp08}, and the first two runs from the
CoRoT space mission (\cite{b06}).

Further data sets that are coming soon to an NStED near you include WASP0 (PI:
S. R. Kane), VULCAN (PI: N. Batalha), BOKS (PIs: S. Howell \&
J. J. Feldmeier), EXPLORE/OC (PIs: K. von Braun \& B. L. Lee), as well as
future CoRoT fields, as NStED is collaborating with the CoRoT team to provide
a NASA portal to the public CoRoT data.

Each featured data set is graciously donated by the respective survey
team. To increase the functionality and usefulness of ETSS, we are soliciting
photometric survey data sets. The NStED webpage
{\tt{http://nsted.ipac.caltech.edu}} provides options to initiate the process
of donating.

\section{Summary}\label{summary}

The NStED Exoplanet Transit Survey Service aims to make time-series photometry
data (light curves) available to the astronomical community in a homogeneous
format. The principal goals are to increase usefulness of the survey data sets
by enabling the extraction of additional scientific results from the
data. ETSS is designed to be straightforward to use and clearly documented.
It is continuously updated to reflect the ingestion of newly donated data sets
as well as the ongoing development of tools to analyze and manipulate the data
contained in the archives. NStED-ETSS is accessible via the NStED homepage at
{\tt{http://nsted.ipac.caltech.edu}}.



\end{document}